\documentclass[superscriptaddress,showpacs,preprintnumbers,amsmath,amssymb,floatfix,prd,fleqn,usenatbib]{mnras}

\usepackage{graphicx}
\usepackage{dcolumn}
\usepackage{bm}
\usepackage{hyperref}
\usepackage{url}
\usepackage{amsfonts}
\usepackage{amsmath}
\usepackage{amssymb}
\usepackage{mathrsfs}
\usepackage{rsfso}
\usepackage{caption}
\usepackage{subcaption}
\usepackage{xcolor}
\definecolor{PURPLE}{RGB}{200,0,120}
\usepackage{float}

\hypersetup{colorlinks,citecolor=blue}
\usepackage{float}
\usepackage{booktabs}
\DeclareRobustCommand{\VAN}[3]{#2}
\let\VANthebibliography\thebibliography
\def\thebibliography{\DeclareRobustCommand{\VAN}[3]{##3}\VANthebibliography}

\title[NGC 7331]{Decoding the Gravitational Geometry and Stellar Photometric Profiles of Galaxy NGC 7331}

\author[A. Sanyal, B. S. Choudhury \& F. Rahaman]{
  Aritra Sanyal$^{1}$\thanks{E-mail: aritrasanyal1@gmail.com},
  Bikramarka S. Choudhury$^{1}$\thanks{E-mail: bikramarka@gmail.com},
  Farook Rahaman$^{1}$\thanks{E-mail: rahaman@associates.iucaa.in}
\\
$^{1}$Department of Mathematics, Jadavpur University, Kolkata 700 032, India
}

\begin{document}

\maketitle

\maketitle
\begin{abstract}
Galaxy rotation curves provide a powerful probe of mass distributions in spiral galaxies. We present a general relativistic analysis of NGC 7331 using its observed rotation curve and WISE W1 (3.4 $\mu$m) photometry to constrain the stellar mass profile. Adopting a static, spherically symmetric spacetime with anisotropic matter (vanishing radial pressure), we fit a modified exponential azimuthal velocity law to kinematic data, reconstructing metric functions and deriving enclosed mass, energy density, and tangential pressure profiles. The stellar mass underpredicts the total gravitating mass at intermediate-to-large radii, indicating dominant dark matter. The physical viability of the resulting relativistic model is examined through energy conditions, causality constraints, and the stability of circular orbits. A comparison with the standard Navarro--Frenk--White dark matter profile has been done. Overall, this study demonstrates that combining rotation-curve data with photometric stellar mass estimates within a general relativistic framework provides a consistent and physically viable description of the mass distribution in spiral galaxies such as NGC~7331.
\end{abstract}

\section{Introduction}

{Galaxies are gravitationally bound systems of stars, gas, dust, and dark matter.
Observationally, matter is not homogeneously distributed throughout the Universe
but instead remains confined within self-contained structures spanning a wide
range of morphologies and spatial scales. Based on their global properties,
galaxies are commonly classified as spiral, elliptical, lenticular
\citep{van2009lenticular}, irregular, or starburst systems \citep{starburst}.
Understanding the physical mechanisms that maintain these bound structures is a
central problem in galactic astrophysics and is closely related to the
distribution of dark matter within galactic halos \citep{matter,halo}.}

{General relativity provides a fundamental description of gravitation,
interpreting gravity as the curvature of spacetime produced by matter and
energy. In this framework, the Einstein field equations relate the spacetime
geometry to the energy--momentum tensor of the matter distribution. For extended
astrophysical systems such as galaxies, static and spherically symmetric
spacetime metrics offer a tractable and widely adopted approximation for
investigating gravitational structure
\citep{Tolman1939,solution1,solution2,solution3}. Such models have been
extensively employed to study relativistic effects in galactic and halo
dynamics.}

{Galaxy rotation curves constitute one of the most powerful observational probes
of galactic mass distribution. Derived from Doppler shifts of optical and radio
emission lines across galactic disks, these curves show that the rotational
velocity of stars and gas remains approximately constant at large
galactocentric radii. This behavior departs significantly from the Keplerian
decline expected from Newtonian dynamics applied to visible matter alone and
provides strong evidence for the presence of an extended, massive dark matter
halo \citep{honma1997rotation,Rubin1980,Sofue2001}.}

{A promising theoretical approach for interpreting galactic rotation profiles is
to analyze them within a general relativistic framework. In particular, static
and spherically symmetric spacetimes combined with anisotropic matter
distributions—especially configurations with vanishing radial pressure—allow a
consistent reconstruction of the underlying gravitational geometry
\citep{Bowers1974,Rahaman2014}. Within this perspective, the observed rotation
curve is not merely a kinematic quantity but directly encodes information about
spacetime curvature through the Einstein field equations \citep{Begeman1989}.}

{A substantial body of literature has explored dark matter models, galactic
dynamics, and observational constraints using both Newtonian and relativistic
approaches \citep{galdark1,galdark2,galdark3}. In this work, we focus on the
well-studied spiral galaxy NGC~7331, whose rotation curve has been measured with
high precision across a wide radial range
\citep{ngc1,ngc2,ngc3,ngc4}. This makes NGC~7331 an ideal laboratory for testing
relativistic mass modeling of galactic halos.}
{
The structure of the present paper is as follows. Section~\ref{sec:photometry}
derives the stellar mass profile of NGC~7331 from near-infrared photometry using
WISE W1 imaging data. Section~\ref{sec:theory} introduces the relativistic
framework and Einstein field equations employed in this study. 
Section~\ref{sec:fitting} describes the galaxy rotation curve data and the
fitting methodology adopted for modeling the azimuthal velocity profile.
Section~\ref{sec:statistics} presents the statistical validation of the fitted
model. Section~\ref{sec:darkmatter} compares the photometric stellar mass
distribution with the total gravitating mass inferred from rotation curve
dynamics, thereby constraining the dark matter distribution. 
Section~\ref{sec:physical} discusses the derived mass, energy density, and
pressure profiles, along with tests of energy conditions and causality.
Section~\ref{sec:stability} examines the stability of circular orbits.
Section~\ref{sec:binding} analyzes the gravitational binding and attractive
nature of the galactic halo. Section~\ref{sec:observational} presents the
effective equation-of-state parameter within the relativistic framework.
Section~\ref{sec:comparison} compares the results with previous studies of
NGC~7331 and with the standard Navarro--Frenk--White (NFW) dark matter profile.
Finally, Section~\ref{sec:conclusion} summarizes the main conclusions and
implications of this work.
}

\section{Derivation of Stellar Mass Profile from WISE Photometry}
\label{sec:photometry}
\begin{figure}
    \centering
    \includegraphics[width=\linewidth]{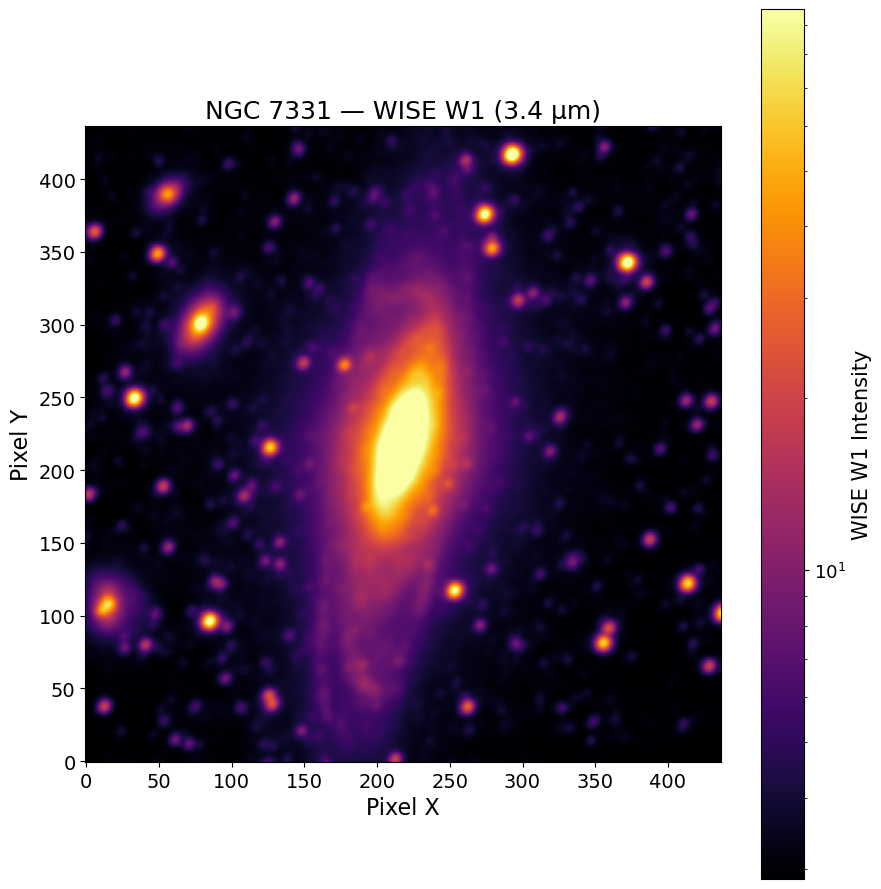}
    \caption{WISE W1 ($3.4\,\mu$m) image of the spiral galaxy NGC~7331 obtained from the NASA/IPAC Infrared Science Archive.}
    \label{fig:wise_w1_ngc7331}
\end{figure}

The stellar mass distribution of the spiral galaxy NGC~7331 is derived using infrared photometric observations obtained by the
\textit{Wide-field Infrared Survey Explorer}  Figure~\ref{fig:wise_w1_ngc7331}(WISE) satellite. shows the WISE W1 ($3.4\,\mu$m) image of
NGC~7331 used to extract the surface brightness profile.
In particular, we utilize data from the WISE W1 band at a wavelength of $3.4~\mu\mathrm{m}$ \citep{IRSAWISE}, which is widely regarded as a robust tracer of stellar mass. Emission at this wavelength is dominated by old,low-mass stars that constitute the bulk of a galaxy’s stellar mass and thus provides a reliable proxy for the underlying stellar mass distribution \citep{Meidt2014,Leroy2019} . Moreover, infrared observations are minimally affected by interstellar dust extinction, allowing an accuratere construction of the stellar distribution even in optically obscured regions.
The observational procedure begins by extracting the surface brightness profile of NGC~7331 from the WISE W1 image, measured as a function of projected galactocentric radius. These brightness measurements are converted into luminosity using standard photometric calibration procedures, while accounting for the distance to NGC~7331, taken to be approximately $14.7~\mathrm{Mpc}$  \citep{Bottema2002}. The luminosity profile is then transformed into a stellar mass profile through the application of an infrared mass-to-light ratio. For the W1 band, a nearly constant mass-to-light ratio in the range
\begin{equation}
\Upsilon_{3.4\,\mu\mathrm{m}} \simeq 0.5 - 0.6~M_\odot/L_\odot
\end{equation}
is commonly adopted, reflecting the relatively uniform contribution of evolved stellar populations at this wavelength\citep{Meidt2014,Leroy2019}.

Using this conversion, the stellar mass enclosed within a given radius is obtained by radially integrating the surface mass density,yielding a discrete set of data points $M_\star(r_i)$,
which represent the cumulative stellar mass as a function of radius. While these discrete measurements accurately represent the observational data, theoretical modeling and dynamical analyses require a smooth, continuous functional form for $M_\star(r)$. Consequently, an analytical fitting procedure is necessary.
Given the high inclination of NGC 7331, projection effects must be taken into account when deriving radial surface brightness and stellar mass profiles.
{
Following standard surface photometry techniques, projection effects due to the high inclination of NGC~7331 are accounted for by extracting the surface brightness profile along elliptical annuli rather than circular ones. The axial ratio of the ellipses is defined as $q = \cos i$, where $i$ denotes the inclination angle of the galactic disk. For NGC~7331, we adopt an inclination of $i = 76^\circ$, consistent with values inferred from photometric axial ratios and kinematic modeling in previous studies
(\citet{deVaucouleurs1991}; \citet{Sofue1997}). The position angle of the major axis is fixed at $\mathrm{PA} = 168^\circ$. The semi-major axis of each ellipse is taken as the effective galactocentric radius, ensuring that the resulting surface brightness profile accurately represents the intrinsic radial distribution in the plane of the galactic disk.}

Several functional forms were tested in order to model the observed stellar mass distribution. These include a simple exponential disk profile, a Sérsic profile, a Plummer model, and a double exponential bulge--disk decomposition. In addition, a modified exponential model was considered, incorporating a power-law behavior in the central region combined with an exponential saturation at large radii. Each model was fitted to the observational data using a least-squares minimization technique.

{Among all tested models, the modified exponential form provided a significantly
superior fit, with residual errors nearly two orders of magnitude smaller than those obtained
from the other profiles. The best-fit parameters are
$M_0 = 1.1271$, $R_d = 26.121~\mathrm{kpc}$, and $\alpha = 1.1008$.
The normalization is chosen such that the stellar mass is unity at the outermost observed radius.
The relatively large scale length indicates an extended stellar disk, while the value of
$\alpha > 1$ signifies a moderately enhanced central mass concentration, consistent with the
presence of a prominent bulge component in NGC~7331.}

\begin{equation}
{
M_\star(r) =
1.1271 \left( \frac{r}{26.121} \right)^{1.1008}
\left( 1 - e^{-r/26.121} \right),
}
\end{equation}

{where $r$ is measured in kiloparsecs. This smooth functional form is subsequently
employed in the dynamical analysis and rotation curve modeling presented in the following sections.}

\begin{figure}
    \centering
    \includegraphics[width=1\linewidth]{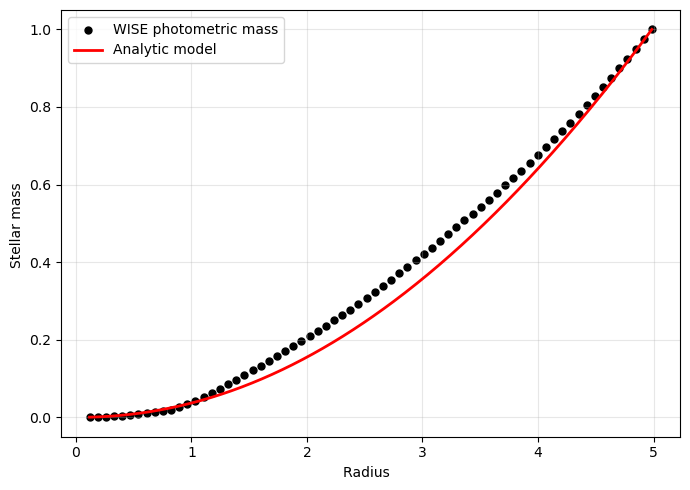}
    \caption{
Stellar mass profile of NGC~7331 derived from WISE W1 photometry.
Black points show the photometric stellar mass, while the red
curve represents the analytic model
}
    \label{fig:placeholder}
\end{figure}

\section{Theoretical Modeling} \label{sec:theory}

\subsection{Spacetime Structure and Field Equations}

We consider a static, spherically symmetric spacetime described by the general line element:
\begin{equation}
    ds^2 = -f(r)\, dt^2 + \frac{dr^2}{1 - \dfrac{2 m(r)}{r}} + r^2 d\Omega^2,
\end{equation}
where \( f(r) \) is the redshift function and \( m(r) \) is the mass function representing the total mass enclosed within radius \( r \). 

To model the matter content generating this geometry, we assume an anisotropic fluid distribution with vanishing radial pressure. The corresponding energy-momentum tensor takes the form:
\begin{equation}
    T^\mu_{\ \nu} = \mathrm{diag}[-\rho(r),\, 0,\, p(r),\, p(r)],
\end{equation}
where \( \rho(r) \) is the energy density and \( p(r) \) denotes the tangential pressure. This configuration is motivated by the interpretation of galactic dark matter as a pressureless dust in the radial direction, with only non-zero tangential (or effective) pressure.

Substituting the above energy-momentum tensor into Einstein's field equations yields the following system:

\begin{itemize}
    \item From the \( G^0_0 \) component, we obtain the differential equation for the mass function:
    \begin{equation} \label{eq:density}
        m'(r) = 4\pi r^2 \rho(r),
    \end{equation}
    which describes how mass accumulates with increasing radius based on the local energy density.
    
    \item The \( G^1_1 \) component provides the evolution of the redshift function in terms of the enclosed mass:
    \begin{equation} \label{eq:mass}
        \frac{f'(r)}{f(r)} = \frac{2m(r)}{r^2 - 2r\, m(r)},
    \end{equation}
    revealing the influence of gravitational geometry on the redshift factor.
    
    \item The \( G^2_2 = G^3_3 \) components yield the expression for the tangential pressure:
    \begin{equation} \label{eq:pressure}
        p(r) = \frac{r\, \rho(r)}{2} \cdot \frac{m(r)}{r^2 - 2r\, m(r)},
    \end{equation}
    thereby closing the system of equations in terms of \( \rho(r) \), \( p(r) \), and \( m(r) \).
\end{itemize}

Together, equations~\eqref{eq:density}–\eqref{eq:pressure} form the backbone of our model. Once a physically motivated form of the redshift function \( f(r) \) is prescribed—either phenomenologically or constrained through observational inputs—this system can be solved to study the distribution and dynamics of matter within galactic halos. The behavior of these functions further allows assessments of energy conditions, stability, and equivalence with observed rotation curves.

{It is worth noting that all calculations in this work have been performed in geometric units. Consequently, the dimensional representations of various physical quantities differ from their standard SI counterparts. In this system, fundamental quantities such as mass, length, time, and energy are expressed in terms of a single unit of length. These conversions follow well-established conventions commonly used in general relativity (see, for example, \cite{Misner}).}

\section{Rotation Curve Modeling and Fitting Methodology}
\label{sec:rotation_fitting}

\subsection{Fitting Methodology}  \label{sec:fitting}

\begin{figure}
    \centering
    \includegraphics[width=\linewidth]{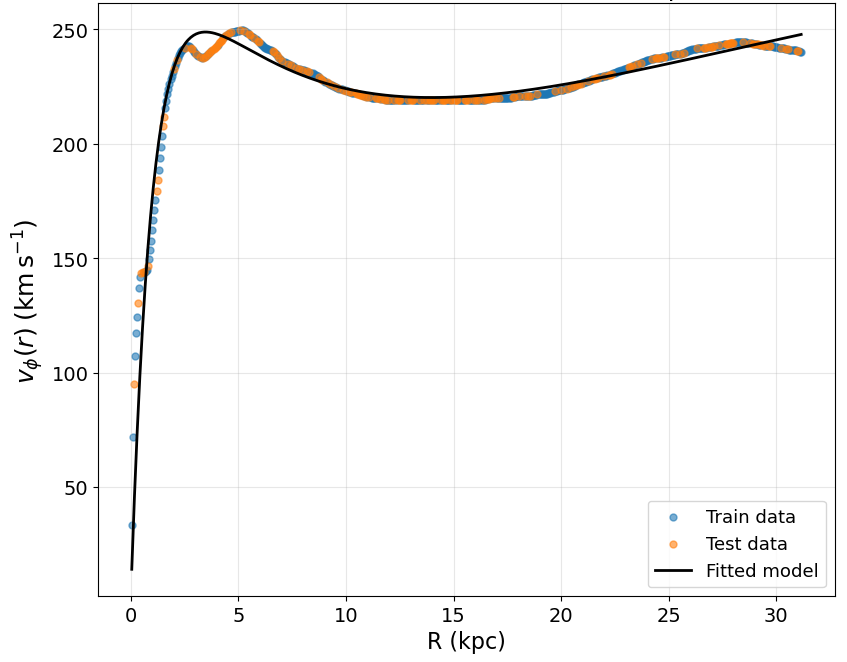}
    \caption{Best-fit rotation curve of NGC~7331 obtained using the modified exponential velocity model. The data are divided using a 70--30 random split into training and testing sets.}
    \label{fig:rotation_split}
\end{figure}

{
The rotation curve of the spiral galaxy NGC~7331 was modeled using a modified exponential azimuthal velocity profile, with parameters \(a\), \(b\), \(c\), and \(d\) determined through nonlinear least-squares fitting to the observed kinematic data shown in Figure~\ref{fig:rotation_split}. The observational data used in this study were obtained from \citet{sofue1999centralrotation}.
The rotation curve compiled combines multiple observational tracers, including optical emission-line spectroscopy (primarily H$\alpha$ and [N\,II] lines from H\,II regions), CO molecular-line observations, and H\,I 21\,cm measurements in the outer disk. This multi-tracer dataset provides reliable kinematic constraints across the full radial extent of NGC~7331. Optical emission lines trace ionized gas in star-forming regions and provide high spatial resolution in the inner galaxy, while radio observations of molecular and atomic gas allow accurate measurements in the outer disk.
To ensure the robustness of the velocity model, several data segmentation strategies were explored, including random, stratified, and sequential splits. The final statistical validation of the model and the selection of the optimal data split are discussed in Section~\ref{sec:statistics}.
}

\subsection{Statistical Analysis} \label{sec:statistics}

{
After evaluating multiple data segmentation strategies, the standard random split method was identified as the most reliable configuration. In this approach, 70\% of the cleaned galaxy rotation data were used for training and the remaining 30\% were reserved for testing. This split provides a well-distributed representation of the dataset across the full galactocentric radial range \(R \in [0.05,\,31.15]\) kpc, corresponding to rotation velocities between 33.42 km\,s\(^{-1}\) and 249.59 km\,s\(^{-1}\).
The fitted model demonstrates strong statistical performance \citep{Bevington1993,Hastie2009}. The coefficient of determination is \(R^2_{\mathrm{train}} = 0.9062\) for the training set and \(R^2_{\mathrm{test}} = 0.8807\) for the test set. The root mean squared error (RMSE) is 6.83 km\,s\(^{-1}\) for the training data and 7.18 km\,s\(^{-1}\) for the test data, indicating minimal overfitting and good generalization capability.Several diagnostic tests were performed to verify the statistical consistency of the model. The reduced chi-square statistic is \(\chi_\nu^2 = 1.0112\) with a corresponding \(p\)-value of 0.4262, indicating good agreement between the model and the observational data. The Durbin--Watson statistic is 2.01, suggesting the absence of significant autocorrelation in the residuals, while the runs test yields a \(p\)-value of 0.1740, confirming that the residuals are randomly distributed.Alternative segmentation strategies were also explored. The \textit{Even--Odd Index Split} produced a test \(R^2\) value of 0.8973 with an RMSE ratio of 1.482, while the \textit{Radial Distance Split} yielded a test \(R^2\) value of 0.8447 with an RMSE ratio of 1.387. Both methods showed weaker predictive performance compared with the standard random split.Overall, the 70--30 random split provides the most balanced and statistically robust configuration. Its strong predictive accuracy and stable residual behaviour make it well suited for modeling galaxy rotation curves and constraining the mass distribution within galactic halos.
}

{\subsection{Comparison with Previous Rotation Curve Studies}
The rotation curve obtained using the present method is in good overall agreement with previously published observational and modeling studies of NGC~7331. In particular, the approximately flat behavior at large galactocentric radii is consistent with the mass modeling results of Flores et al.~\cite{flores1993}. The overall amplitude and radial extent of the fitted curve also compare well with the more recent analysis of Ghari et al.~\cite{ghari2019}, who performed a detailed decomposition of baryonic and dark matter components.Differences are mainly confined to the inner region ($r \lesssim 5$\,kpc), where earlier studies explicitly incorporate bulge--disk--halo decompositions, whereas the present approach employs a smooth relativistic velocity profile without component-wise separation. A localized deviation near $r \simeq 4$\,kpc, associated with the well-known circumnuclear ring of NGC~7331, is likewise consistent with features discussed in previous works and reflects physical structures not captured within a purely axisymmetric model. Similar sensitivities of the inner rotation curve to modeling assumptions and tracer selection have also been emphasized by Sylos Labini et al.~\cite{labini2023}. Overall, the level of agreement with earlier studies supports the robustness of the rotation curve produced by the method presented here.}

\subsection{Rotation Curve Data and Velocity Model}

The rotation curve of NGC~7331 is modeled using a modified exponential azimuthal velocity profile fitted to observational data from Sofue et al.~\cite{sofue1999centralrotation}, derived from optical (H$\alpha$, [N\,\textsc{ii}]) and radio (CO and H\,\textsc{i} 21\,cm) observations. A standard 70\%--30\% random split of the data is adopted to ensure balanced radial coverage.

The azimuthal velocity profile is given by
\begin{equation}\label{eq:velocity_model}
v_\phi(r) = a r e^{-br} + c \left(1 - e^{-dr}\right).
\end{equation}

The best-fit parameters obtained from the nonlinear least-squares fitting are listed in Table~\ref{tab:velocity_parameters}.

\begin{table}
\centering
\fbox{
\begin{tabular}{lc}
\hline
Parameter & Best-fit value \\
\hline
$a$ & $-15.4409 \pm 0.6004$ \\
$b$ & $0.0715 \pm 0.0008$ \\
$c$ & $299.6312 \pm 2.8439$ \\
$d$ & $1.0105 \pm 0.0205$ \\
\hline
\end{tabular}
}
\caption{Best-fit parameters of the modified exponential velocity model for NGC~7331.}
\label{tab:velocity_parameters}
\end{table}

\subsection{Computing the Redshift Function from Observed Velocity Data}  \label{sec:redshift}

To construct the redshift function $f(r)$, we begin with the
relativistic relation connecting it to the squared azimuthal
velocity of test particles in circular motion,
\begin{equation}
\frac{df}{dr} = \frac{v_\phi^2(r)}{r}.
\label{eq:redshift_diff}
\end{equation}
\begingroup
{Using the fitted azimuthal velocity profile given in Eq.~\eqref{eq:velocity_model}}, we substitute directly into Eq.~(\ref{eq:redshift_diff}) to obtain
\begin{equation}
\frac{df}{dr}
= \frac{1}{r}
\left[
a r e^{-br} + c \left(1 - e^{-dr}\right)
\right]^2 .
\end{equation}
\endgroup
Integrating with respect to $r$, the redshift function is then
given by
\begin{equation}
f(r) = \int \frac{v_\phi^2(r)}{r}\,dr .
\end{equation}
Carrying out the integration analytically yields
\begin{align}\label{redshift_func}
f(r) &= -0.143\, r
- 1667.282\, r\, e^{-0.143 r}
+ 89778.856 \log r \nonumber \\
&\quad + 89778.856\, \mathrm{Ei}(-2.021 r)
- 179557.712\, \mathrm{Ei}(-1.0105 r) \nonumber \\
&\quad - 8551.895\, e^{-1.082 r}
- 11659.318\, e^{-0.143 r} \nonumber \\
&\quad + 129414.696\, e^{-0.0715 r},
\end{align}
where $\mathrm{Ei}(-x)$ denotes the exponential integral
function defined as
\begin{equation}
\mathrm{Ei}(-x) = - \int_x^{\infty} \frac{e^{-t}}{t}\,dt .
\end{equation}

\section{Dark Matter Distribution in NGC~7331}
\label{sec:darkmatter}
{The total gravitating mass distribution of NGC~7331 used in this analysis is obtained directly from the relativistic framework introduced in Section~3. After reconstructing the redshift function $f(r)$ from the fitted azimuthal velocity profile (Section~4.5), the enclosed mass function $m(r)$ is determined from the Einstein field equation~(6). This procedure follows the observationally guided redshift-function methodology developed in \citet{Rahaman2026}, where the spacetime geometry is reconstructed consistently from statistically validated rotation-curve data.}

In contrast, the stellar mass profile obtained from WISE W1 photometry traces only the luminous baryonic component associated with the old stellar population. As shown in the Section~2, the enclosed stellar mass increases rapidly in the inner regions of the galaxy but approaches saturation at larger radii, reflecting the finite extent of the stellar disk.

A comparison between the total gravitating mass profile and the stellar mass distribution reveals a systematic excess of mass at intermediate and large radii, where the stellar contribution alone is insufficient to account for the observed rotation velocities. This excess mass can be quantified by defining the dark matter mass profile as
\begin{equation}
M_{\rm DM}(r) = M_{\rm total}(r) - M_\star(r),
\end{equation}
where $M_{\rm total}(r)$ is the mass inferred from the rotation curve and $M_\star(r)$ is the stellar mass derived from photometry shown in figure ~\ref{fig:placeholder}.

The resulting dark matter distribution dominates the mass budget beyond the optical extent of the galaxy, indicating that a substantial non-luminous component is required to explain the observed dynamics of NGC~7331. In the inner regions, the stellar mass contributes significantly to the gravitational potential, while at larger radii the dark matter component becomes increasingly dominant. This transition from baryon-dominated to dark-matter-dominated dynamics is a well-established characteristic of spiral galaxies and is clearly reproduced in the present analysis.

The inferred dark matter profile provides an essential observational constraint for subsequent comparisons with standard halo models and for assessing the consistency of the relativistic mass modelling framework adopted in this work.

\begin{figure}
    \centering
    \includegraphics[width=\linewidth]{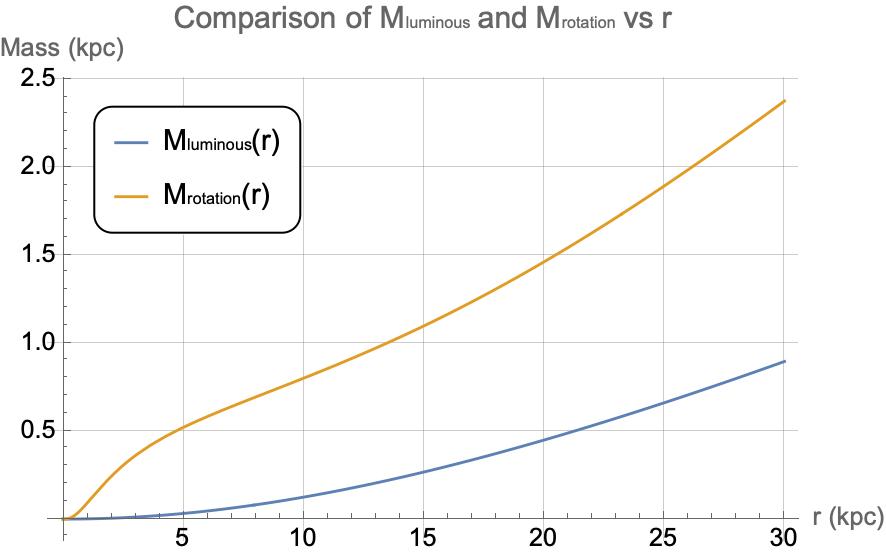}
    \caption{$M_{luminous}$ denotes the mass inferred from photometric data whereas $M_{rotational}$ denotes the mass derived from the rotational curve of the galaxy.}
    \label{fig:placeholder}
\end{figure}

\section{Physical Analysis} \label{sec:physical}

\subsection{Mass, Pressure, and Density}

\begin{figure}
    \centering
    \includegraphics[width=0.9\linewidth]{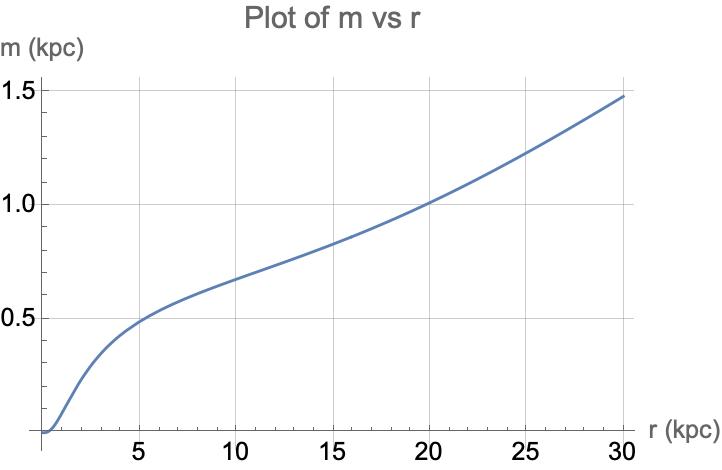}
    \caption{Plot of $m(r)=M_{rotational}(r) - M_{luminous}(r)$ vs $r$ for radial values ranging from 0.1$kpc$ to 30$kpc$.}
    \label{fig:mass}
\end{figure}

\begin{figure}
  \centering
  \begin{subfigure}[b]{0.49\textwidth}
    \centering
    \includegraphics[width=0.9\textwidth]{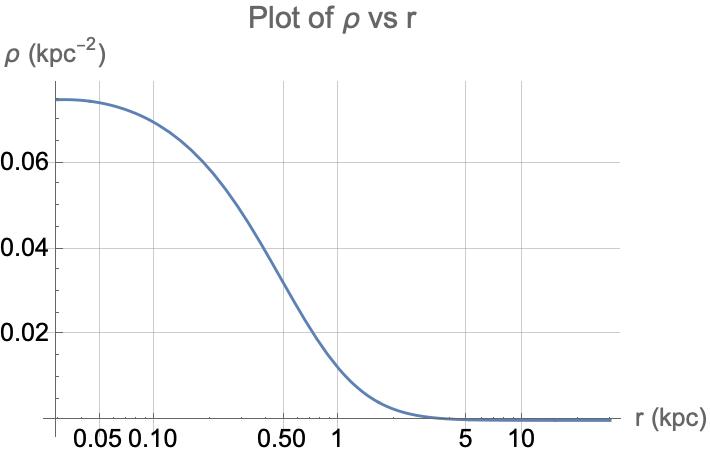}
    \caption{}
    \label{fig:sub1}
  \end{subfigure}
  \hfill
  \begin{subfigure}[b]{0.49\textwidth}
    \centering
    \includegraphics[width=0.9\textwidth]{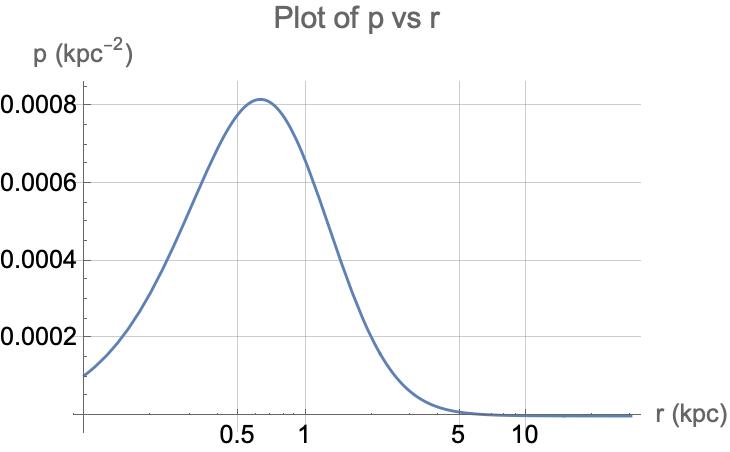}
    \caption{}
    \label{fig:sub2}
  \end{subfigure}

\caption{Plot of (a) $\rho$ vs $r$, (b) $p$ vs $r$, for radial values ranging from 0.1$kpc$ to 30$kpc$}
\label{fig:den_press}
\end{figure}

Now we calculate the expressions of mass, pressure and density for the galaxy through our exponential approximation of azimuthal velocity. For this we will use the field equations \eqref{eq:density}-\eqref{eq:pressure} and our approximation of redshift function given in Section \ref{redshift_func}.

From equation \eqref{eq:mass}, we get
\begin{equation}
    m(r) =\frac{1}{2}. \frac{r^2 f'(r)}{f(r)+rf'(r)}
\end{equation}
The expression for $f(r)$ is given by \eqref{redshift_func}. Therefore from the above equation, we obtain the form of $m(r)$ which is mass obtained from the rotational curve data. {By subtracting the radial mass distribution obtained from the photometric data, we can isolate the mass distribution of matter that is electromagnetically non-interactive but has gravitational effects, i.e. the exotic matter.  For the rest of our analysis, we use only this mass to obtain expressions for the density and pressure profiles of the exotic matter distribution in this galaxy.}

Fig.~\ref{fig:mass} denotes that the exotic mass enclosed within radius $r$ increases as $r$ increases and Fig.~\ref{fig:den_press} expresses the energy density and pressure component gradually falls off as we move away from the center. We see the transverse pressure component has a local maxima at about $0.7kpc$ while the density component does not show any such features.

\subsection{Energy Conditions}

Energy conditions provide essential criteria for assessing the physical viability of relativistic matter distributions in astrophysical systems. They ensure that energy density and pressure behave in a physically reasonable manner and are commonly used to test whether a given spacetime geometry can represent a realistic gravitating system.

For extended galactic halos, smooth and monotonic radial variations of the energy density and pressure are generally indicative of dynamical equilibrium. Conversely, strong irregularities or unphysical sign changes in these quantities may signal departures from equilibrium or the presence of non-standard matter behavior. Energy conditions therefore serve as an important consistency check for relativistic halo models.

The four principal energy conditions considered in this work are given by:
\begin{itemize}
    \item Null Energy Condition (NEC): \hspace{2em} $\rho + p_r \geq 0, \qquad \rho + p_t \geq 0$
    \item Weak Energy Condition (WEC): \hspace{1.2em} $\rho \geq 0, \qquad \rho + p_r \geq 0, \qquad \rho + p_t \geq 0$
    \item Strong Energy Condition (SEC): \hspace{1.2em} $\rho + p_r \geq 0, \qquad \rho + p_t \geq 0, \qquad \rho + p_r + 2p_t \geq 0$
    \item Dominant Energy Condition (DEC): $\rho \geq |p_r|, \qquad \rho \geq |p_t|$
\end{itemize}
where $\rho$ denotes the energy density, $p_r$ the radial pressure, and $p_t$ the tangential pressure.

For the present model, the radial pressure is assumed to vanish ($p_r = 0$), corresponding to an anisotropic matter distribution appropriate for galactic halos. Under this assumption, the energy conditions reduce to:
\begin{itemize}
    \item Null Energy Condition (NEC): \hspace{2em} $\rho + p \geq 0$
    \item Weak Energy Condition (WEC): \hspace{1.2em} $\rho \geq 0, \quad \rho + p \geq 0$
    \item Strong Energy Condition (SEC): \hspace{1.2em} $\rho + p \geq 0, \quad \rho + 2p \geq 0$
    \item Dominant Energy Condition (DEC): $\rho \geq |p|$
\end{itemize}
where $p \equiv p_t$ is the tangential pressure.

\begin{figure}
  \centering
  \begin{subfigure}[b]{0.49\textwidth}
    \centering
    \includegraphics[width=\textwidth]{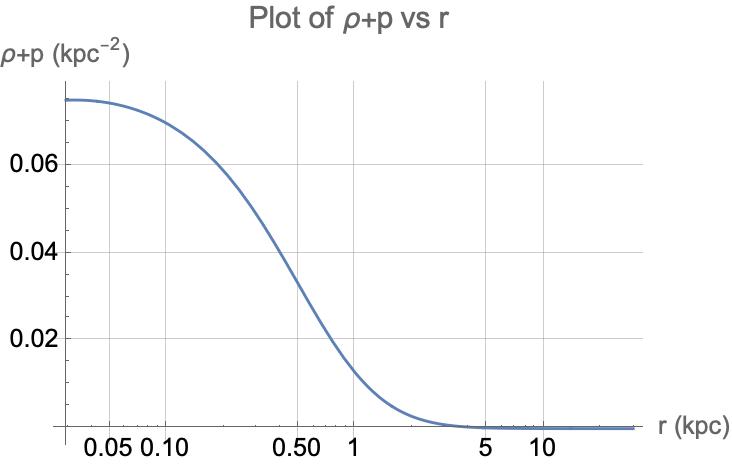}
    \caption{$\rho + p$}
    \label{fig:energy_a}
  \end{subfigure}
  \hfill
  \begin{subfigure}[b]{0.49\textwidth}
    \centering
    \includegraphics[width=\textwidth]{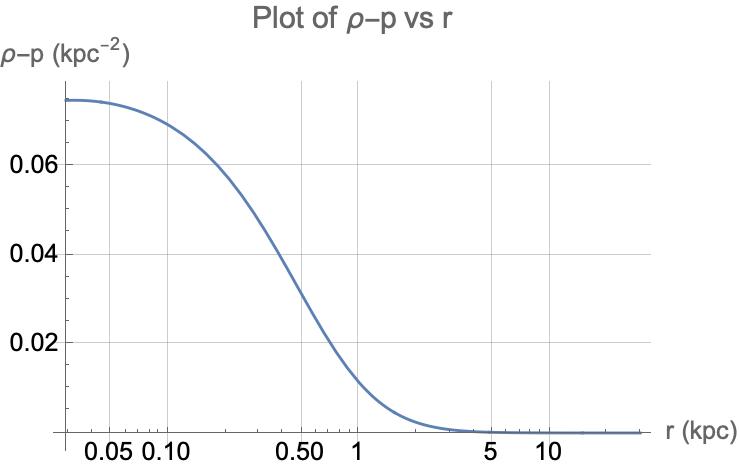}
    \caption{$\rho - p$}
    \label{fig:energy_b}
  \end{subfigure}

  \vspace{0.3cm}

  \begin{subfigure}[b]{0.49\textwidth}
    \centering
    \includegraphics[width=\textwidth]{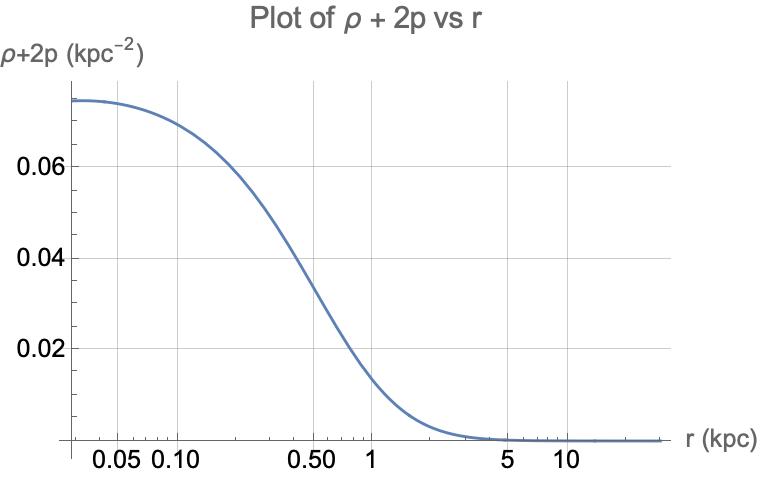}
    \caption{$\rho + 2p$}
    \label{fig:energy_c}
  \end{subfigure}
  \caption{Radial behavior of the energy-condition combinations for $0.1\,\mathrm{kpc} \le r \le 30\,\mathrm{kpc}$. Panels show (a) $\rho + p$, (b) $\rho - p$, and (c) $\rho + 2p$.}
  \label{fig:energy}
\end{figure}

From the derived profiles, both the energy density $\rho(r)$ and the tangential pressure $p(r)$ remain positive throughout the radial range considered. As shown in Fig.~\ref{fig:energy}, all the required inequalities for the Null, Weak, Strong, and Dominant Energy Conditions are satisfied across the halo. This confirms that the relativistic matter distribution supporting the spacetime geometry of NGC~7331 is physically viable within the adopted framework.

\subsection{Pressure--Density Relationship}

The relationship between pressure $p$ and energy density $\rho$ provides a useful diagnostic of the effective matter behavior supporting a gravitating system. In relativistic models of galactic halos, this relation does not represent a thermodynamic equation of state in the traditional sense, but rather reflects how the effective pressure responds to variations in the energy density along the radial direction of the system.

\begin{figure}
    \centering
    \includegraphics[width=\linewidth]{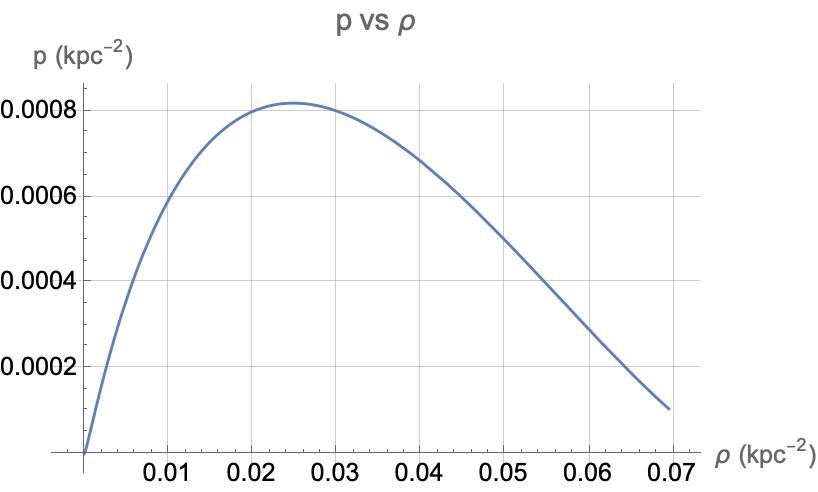}
    \caption{Parametric plot of tangential pressure $p$ versus energy density $\rho$, with the radial coordinate ranging from $0.1\,\mathrm{kpc}$ to $30\,\mathrm{kpc}$.}
    \label{fig:pressure_density}
\end{figure}

Figure~\ref{fig:pressure_density} shows the parametric variation of pressure with density for the exponential velocity model applied to NGC~7331. The relation is clearly non-linear and non-monotonic: the pressure increases with density up to a maximum value and then decreases as the density continues to increase. This behavior indicates that the effective pressure response of the halo matter is not governed by a simple linear relation.

The presence of a pressure maximum suggests that the system exhibits its greatest resistance to compression at a characteristic density. Beyond this point, further increases in density are accompanied by a reduction in pressure, which may reflect changes in the balance between gravitational confinement and anisotropic stresses within the halo. Such behavior is commonly encountered in relativistic models with anisotropic pressure distributions and does not necessarily imply any abrupt change in the underlying matter composition.

Overall, the observed pressure--density relationship highlights the complexity of the effective matter behavior in galactic halos when modeled within a general relativistic framework. It provides additional support for treating the halo as an anisotropic system whose properties vary smoothly with radius, rather than assuming a simple barotropic equation of state.

\subsection{Equation of State Parameter}

The equation of state (EoS) parameter provides a convenient way to characterize the relative contribution of pressure to the energy density in a gravitating system. In the present analysis, we define the effective equation of state parameter as
\begin{equation}
\omega(r) = \frac{p(r)}{\rho(r)},
\end{equation}
where $\rho(r)$ denotes the energy density and $p(r)$ represents the tangential pressure at a radial distance $r$. Since the radial pressure vanishes in our model, the EoS parameter reflects the effective anisotropic pressure contribution supporting the galactic halo.

At galactic scales, the parameter $\omega$ should not be interpreted as a cosmological equation of state. Instead, it serves as a phenomenological measure of how pressure effects compare with energy density within the relativistic halo framework. Small values of $\omega$ correspond to systems dominated by energy density with negligible pressure support, consistent with expectations for dark matter halos.

\begin{figure}
    \centering
    \includegraphics[width=\linewidth]{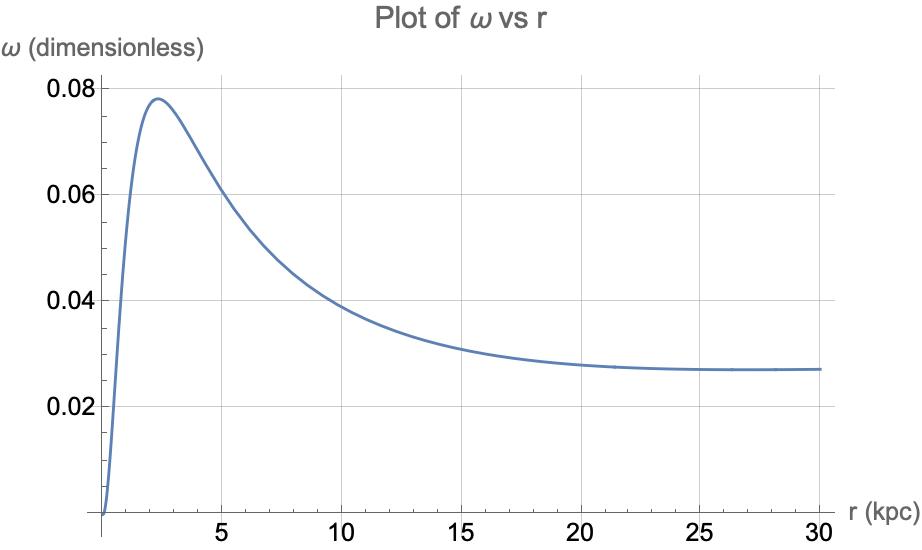}
    \caption{Radial variation of the effective equation of state parameter $\omega(r)$ for $0.1\,\mathrm{kpc} \le r \le 30\,\mathrm{kpc}$.}
    \label{fig:omega}
\end{figure}

Figure~\ref{fig:omega} shows that the EoS parameter remains small and positive throughout the radial range considered, with values typically lying between $0$ and $0.1$. This indicates that the pressure contribution is significantly smaller than the energy density, consistent with a weakly pressured, dark-matter-dominated halo. The smooth radial behavior of $\omega(r)$ further supports the interpretation of the halo matter as an anisotropic but physically stable system within the adopted relativistic framework.

\subsection{Causality (velocity of sound)}

In the study of a galaxy modeled as a stellar object, the causality conditions require that the speed of sound do not exceed the speed of light. These constraints ensure physical viability:
\begin{align}\label{sound_eq}
0 \leq v_{\text{sound}}^2 = \frac{dp}{d\rho} \leq 1,
\end{align}
where $p$ is the pressure, and $\rho$ is the energy density. These inequalities guarantee that signals propagate at subluminal speeds.

\begin{figure}
    \centering
    \includegraphics[width=\linewidth]{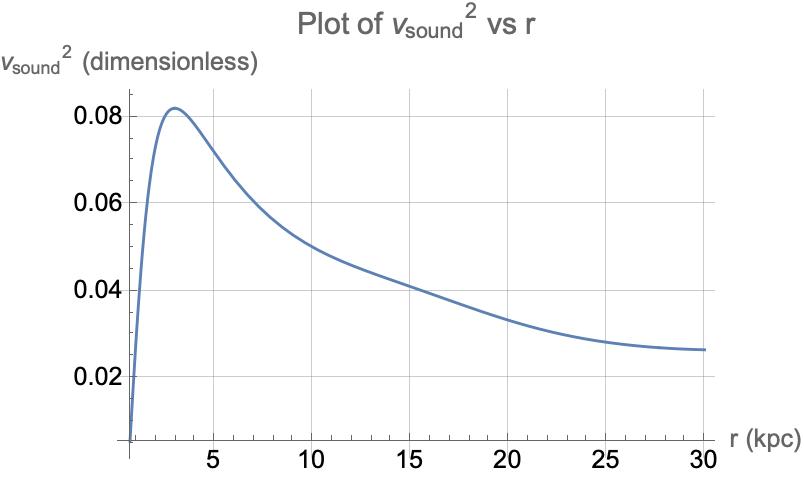}
    \caption{Plot for the squared sound velocity vs $r$ for the radial values ranging from 0.65$kpc$ to 30$kpc$}
    \label{fig:sound}
\end{figure}

\begin{figure*}
    \centering
    \includegraphics[width=0.8\linewidth]{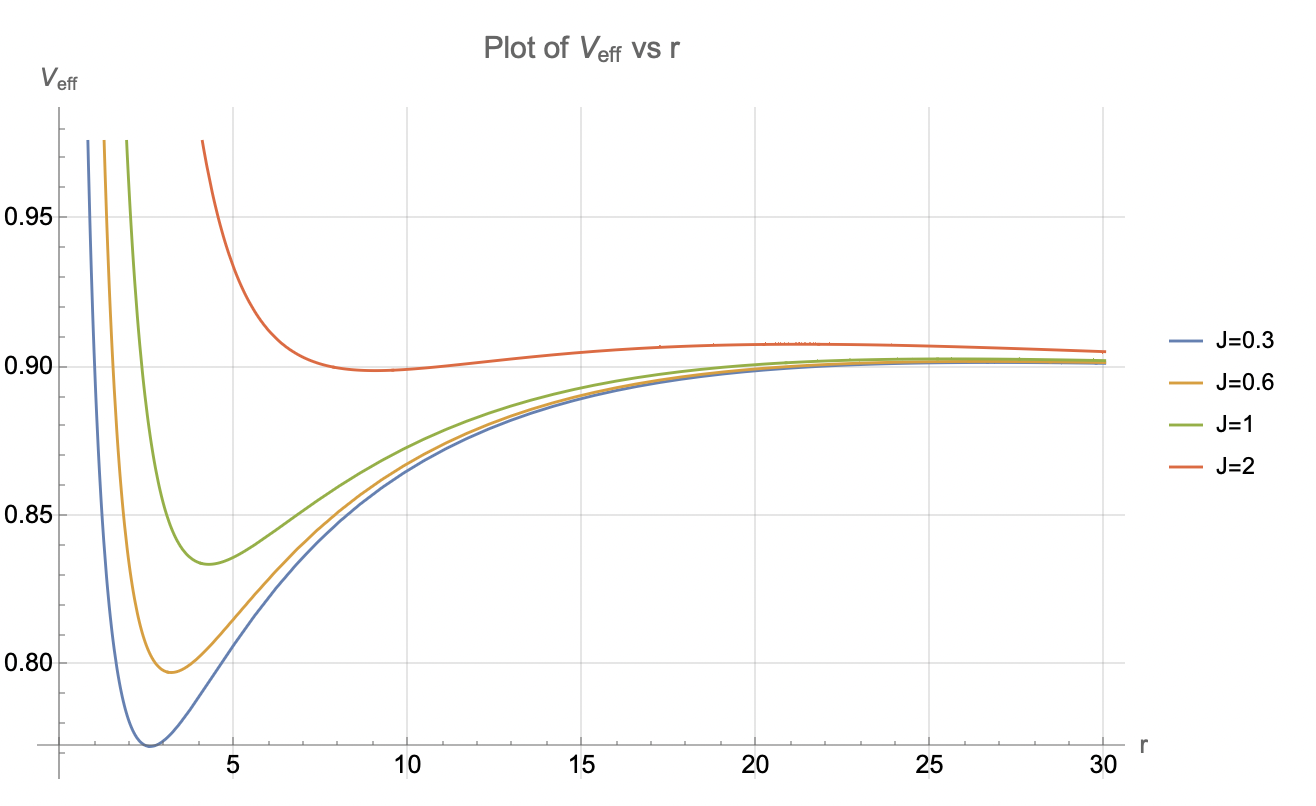}
    \caption{Plot of $V_{\text{eff}}$ vs $r$ for differnet values for angular momentum per unit mass ($J$)}
     \label{fig:potential}
\end{figure*}

From the Fig.~\ref{fig:sound}, we can clearly see the plot of the squared velocity of sound indeed satisfies the constraint given by eq. \eqref{sound_eq}. Hence the causality conditions are also satisfied in our model.

\section{STABILITY OF CIRCULAR ORBITS} \label{sec:stability}

Restricting to equatorial plane motion (\( \theta = \pi/2 \)) and defining the four-velocity \( U^\alpha = \frac{dx^\alpha}{d\tau} \), the normalization condition \( g_{\mu\nu} U^\mu U^\nu = -1 \) leads to a radial equation of the form
\[
\left(\frac{dr}{d\tau}\right)^2 = E^2 - V_{\text{eff}}(r),
\]
where the effective potential is given by
\[
V_{\text{eff}}(r) = \left(1 - \frac{2m(r)}{r} \right)\left(1 + \frac{J^2}{r^2}\right).
\]

Here \( E \) and \( J \) are the conserved relativistic energy and angular momentum per unit rest mass, defined by
\[
E = f(r)\dot{t}, \quad J = r^2\dot{\phi}.
\]

Circular orbits are defined by constant radius \( r = R \), so that
\[
\left. \frac{dr}{d\tau} \right|_{r = R} = 0 \quad \text{and} \quad \left. \frac{dV_{\text{eff}}}{dr} \right|_{r = R} = 0.
\]

Now for the values corresponding to the exponential approximation we plot the potential function for certain angular momentum values in the following plot in Fig.~\ref{fig:potential}.

Here, through the plot of Fig.~\ref{fig:potential}, we see in each case the effective potential has a minima near the central region for different angular momentum values. The lower the angular momentum per unit mass is, the more prominent the minima is. This minima indicates existence of stable circular orbits near that region.

\section{Gravitational Binding and Attractive Nature of the Halo}
\label{sec:binding}

To examine the attractive nature of the spacetime describing the galactic halo, we consider the motion of a test particle initially at rest in the equatorial plane ($\theta = \pi/2$). The particle trajectory is governed by the geodesic equation
\begin{equation}
    \frac{d^2 r}{d\tau^2} + \Gamma^r_{\mu\nu} \frac{dx^\mu}{d\tau} \frac{dx^\nu}{d\tau} = 0,
\end{equation}
where $\tau$ denotes the proper time.

At the point of release, the particle has vanishing radial velocity, $\dot r = 0$. The radial component of the geodesic equation then reduces to
\begin{equation}
    \left.\frac{d^2 r}{d\tau^2}\right|_{r=r_0}
    = -\Gamma^r_{tt}\left(\frac{dt}{d\tau}\right)^2
      -\Gamma^r_{\phi\phi}\left(\frac{d\phi}{d\tau}\right)^2 .
\end{equation}

For the static, spherically symmetric metric
\begin{equation}
    ds^2 = -f(r)\,dt^2 + \frac{dr^2}{1-\dfrac{2m(r)}{r}}
    + r^2\left(d\theta^2+\sin^2\theta\,d\phi^2\right),
\end{equation}
the relevant Christoffel symbols are
\begin{equation}
    \Gamma^r_{tt}
    = \frac{f'(r)}{2}\left(1-\frac{2m(r)}{r}\right),
    \qquad
    \Gamma^r_{\phi\phi}
    = -r\left(1-\frac{2m(r)}{r}\right).
\end{equation}

Substituting these expressions yields
\begin{equation}
\left.\frac{d^2 r}{d\tau^2}\right|_{r=r_0}
= -\left(1-\frac{2m(r_0)}{r_0}\right)
\left[
\frac{f'(r_0)}{2}\left(\frac{dt}{d\tau}\right)^2
- r_0\left(\frac{d\phi}{d\tau}\right)^2
\right].
\end{equation}

While the above expression already encodes the local attractive character of the spacetime, a complementary global assessment can be obtained by evaluating the gravitational energy contained between two radii $r_1$ and $r_2$. Following standard definitions, the total gravitational energy is given by
\begin{equation}
E_G = 4\pi \int_{r_1}^{r_2}
\left[1-\sqrt{g_{rr}(r)}\right]\rho(r)\,r^2\,dr,
\end{equation}
where the radial metric component is
\begin{equation}
g_{rr}(r) = \frac{1}{1-\dfrac{2m(r)}{r}} .
\end{equation}

This leads to
\begin{equation}
E_G = 4\pi \int_{r_1}^{r_2}
\left[1-\left(1-\frac{2m(r)}{r}\right)^{-1/2}\right]
\rho(r)\,r^2\,dr .
\end{equation}

Using the Einstein field equation $G^0{}_0 = 8\pi\rho$, the energy density can be expressed as
\begin{equation}
\rho(r) = \frac{1}{4\pi}\frac{m'(r)}{r^2}.
\end{equation}
Substitution then yields
\begin{equation}
E_G = \int_{r_1}^{r_2}
\left[1-\left(1-\frac{2m(r)}{r}\right)^{-1/2}\right]
m'(r)\,dr .
\label{grav}
\end{equation}

\begin{figure}
    \centering
    \includegraphics[width=0.95\linewidth]{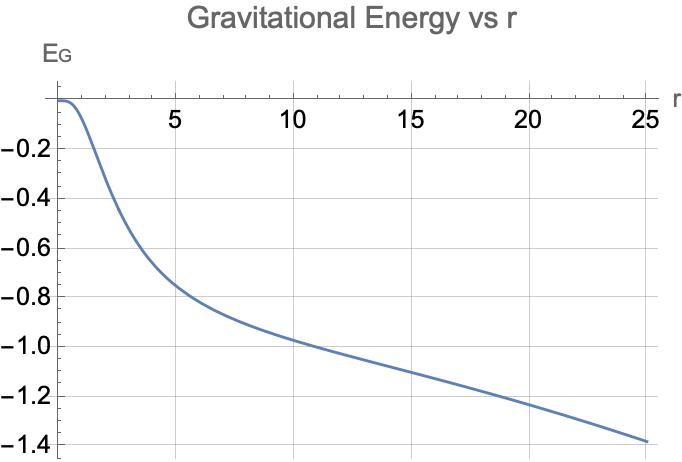}
    \caption{Gravitational energy $E_G$ as a function of radius, computed from Eq.~(\ref{grav}) with the lower limit fixed at $r_1=0.5\,\mathrm{kpc}$. {This shows us the total gravitational energy contained within the radius $0.5\,$kpc and $r\,$kpc}}
    \label{fig:gravitational_energy}
\end{figure}

Figure~\ref{fig:gravitational_energy} shows that the gravitational energy remains negative throughout the radial range considered,
\begin{equation}
E_G < 0 .
\end{equation}
A negative gravitational energy is a clear indication that the spacetime configuration is gravitationally bound and that the effective gravitational interaction within the halo is attractive. This result is consistent with expectations for a physically viable galactic halo model.

\section{Equation of State Parameter}
\label{sec:observational}

For a static, spherically symmetric spacetime of the form
\begin{equation}
ds^{2} = -e^{2\Phi(r)}\,dt^{2}
+ \frac{dr^{2}}{1-\dfrac{2m(r)}{r}}
+ r^{2}\left(d\theta^{2} + \sin^{2}\theta\, d\varphi^{2}\right),
\end{equation}
the redshift function associated with circular motion can be approximated, following \citet{FaberVisser2006}, as
\begin{equation}
z_{\pm}^{2} \approx r\,\Phi'(r).
\label{redshift1}
\end{equation}

For the metric employed in this work, the redshift potential is related to the metric function $f(r)$ through
\begin{equation}
\Phi(r) = \frac{1}{2}\ln f(r).
\end{equation}
Substituting this relation into Eq.~(\ref{redshift1}), we obtain
\begin{equation}
z_{\pm}^{2} \approx \frac{r}{2f(r)}\,f'(r).
\label{redsh}
\end{equation}

\begin{figure}
    \centering
    \includegraphics[width=0.95\linewidth]{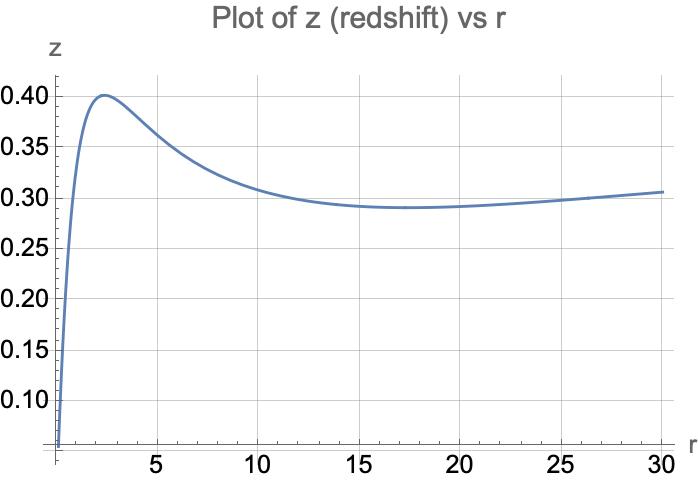}
    \caption{Radial variation of the Faber--Visser redshift function defined in Eq.~(\ref{redsh}) for $0.5\,\mathrm{kpc} \le r \le 25\,\mathrm{kpc}$.}
    \label{fig:redshift}
\end{figure}

Fig.~\ref{fig:redshift} visualizes this redshift function as a function of radius.

Within this relativistic framework, an effective pseudo-mass associated with rotation curves can be defined in terms of the redshift potential as
\begin{equation}
m_{\mathrm{RC}} \equiv r^{2}\Phi'(r).
\end{equation}
This quantity does not represent a conserved global mass but rather a local effective mass that probes the combined influence of spacetime geometry and matter content. In the weak-field limit, it reduces to the post-Newtonian mass expression
\begin{equation}
m_{\mathrm{RC}} \approx 4\pi \int (\rho + p_r + 2p_t)\,r^{2}\,dr \equiv M_{\mathrm{pN}},
\end{equation}
which incorporates contributions from the energy density $\rho$, radial pressure $p_r$, and tangential pressure $p_t$. In the limit of negligible pressure, $m_{\mathrm{RC}}$ coincides with the Newtonian mass profile $M(r)$.

An additional pseudo-mass relevant for lensing observations is given by \citet{FaberVisser2006} as
\begin{equation}
m_{\mathrm{lens}} = \frac{1}{2} r^{2}\Phi'(r) + \frac{1}{2} m(r).
\end{equation}

Using these definitions, \citet{FaberVisser2006} introduced a dimensionless effective equation-of-state parameter,
\begin{equation}
w(r) = \frac{p_r + 2p_t}{3\rho}
\approx
\frac{2}{3}
\frac{m_{\mathrm{RC}}' - m_{\mathrm{lens}}'}
     {2m_{\mathrm{lens}}' - m_{\mathrm{RC}}'},
\label{eos_w}
\end{equation}
which reduces to $w = p/\rho$ in the isotropic-pressure limit.

\begin{figure}
    \centering
    \includegraphics[width=\linewidth]{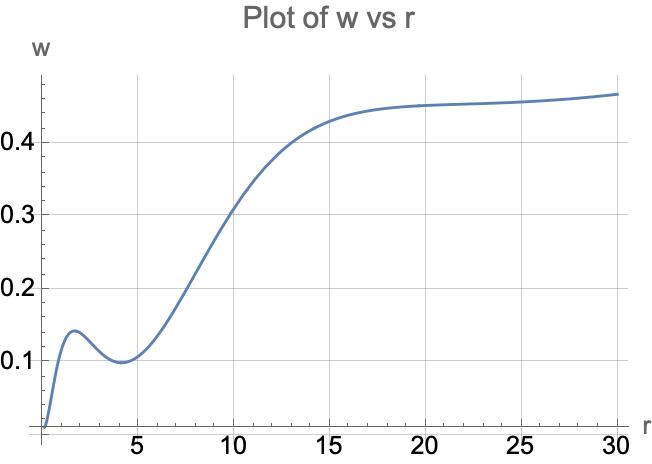}
    \caption{Radial behavior of the Faber--Visser effective equation-of-state parameter $w(r)$ defined in Eq.~(\ref{eos_w}) for the NGC~7331 halo model.}
    \label{fig:eos_fv}
\end{figure}

Figure~\ref{fig:eos_fv} shows the radial variation of $w(r)$ for the exponential velocity model applied to NGC~7331. The parameter exhibits a non-monotonic but bounded behavior across the halo. At galactic scales, this quantity should be interpreted as an effective diagnostic of anisotropic pressure contributions within the relativistic matter distribution rather than as a cosmological equation of state. The observed behavior of $w(r)$ reflects the interplay between geometry and anisotropic stresses in the halo and does not imply the presence of cosmological dark energy.

\section{Comparison with Standard Dark Matter Profiles} \label{sec:comparison}

\begin{figure}
    \centering
    \includegraphics[width=1\linewidth]{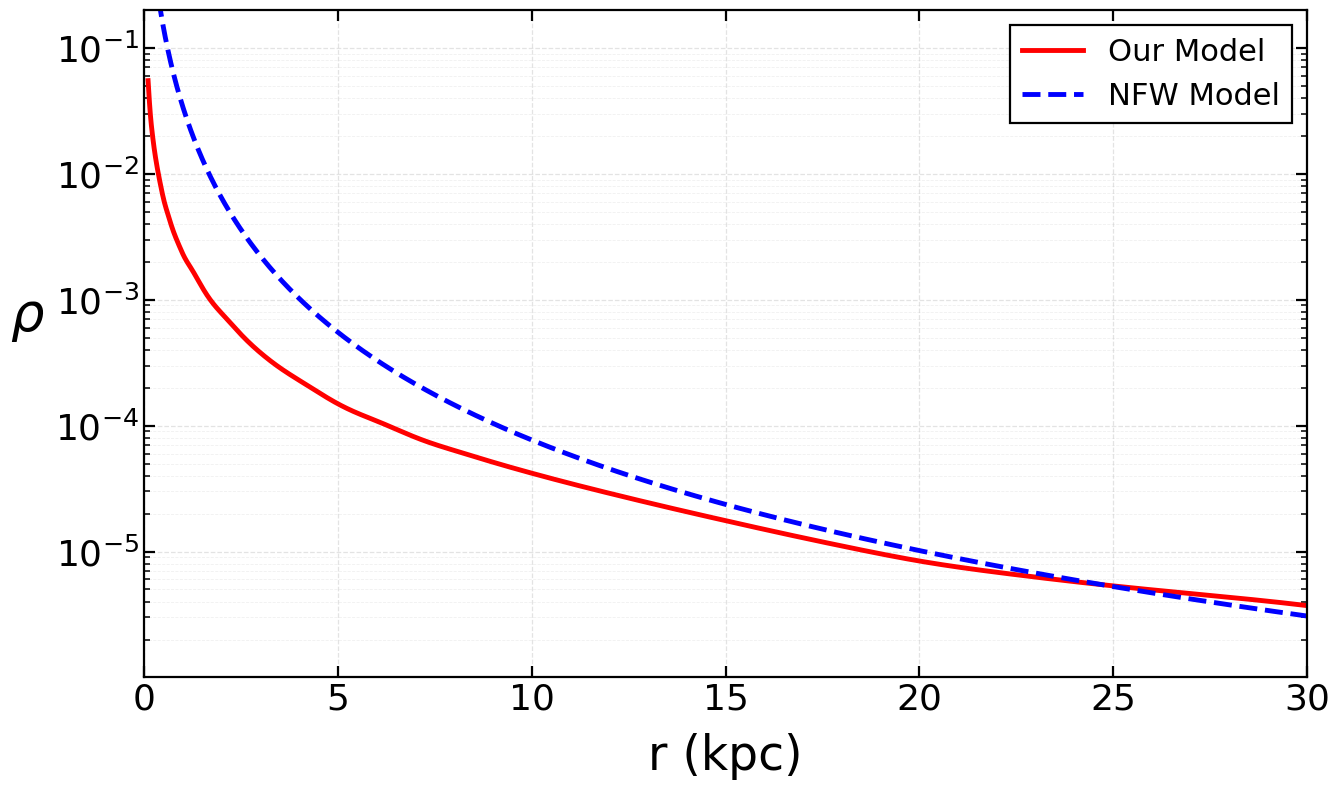}
    \caption{Comparison of dark matter density profiles in NGC 7331 taking the NFW parameter values as $r_N=0.6$}
    \label{fig:NFW}
\end{figure}

{
The plot in Fig.~\ref{fig:NFW}, comparing our density approximation with the standard Navarro-Frenk-White (NFW) profile for the dark matter halo of NGC 7331, reveals several important insights. Both profiles display a smooth, monotonically decreasing density as the distance from the galactic center increases, consistent with expectations for dark matter halos. Our approximation shows notable deviations from the NFW profile across different radial regimes. However, at smaller radii (below 3 kpc), our model predicts a significantly lower central density than the NFW profile, with our densities reaching only 23\% to 41\% of the NFW values, suggesting a less concentrated or flatter inner halo structure compared to the characteristic NFW cusp. At intermediate radii (around 3 to 10 kpc), our approximation continues to fall below the NFW curve, with density ratios ranging from approximately 29\% to 32\%, indicating a more extended mass distribution in this region.

Conversely, at larger radii beyond approximately 15 kpc, our approximation begins to approach and eventually exceed the NFW profile, reaching 79\% at 20 kpc and rising to 148\% at 30 kpc, implying a shallower decline or less rapid truncation of the halo density. The smooth analytical nature of our approximation, derived from the relativistic field equations, contrasts with the universal form of the NFW profile characterized by parameters $r_N = 0.6$ kpc and $\rho_N = 0.4$. Overall, this comparison validates our model's effectiveness in capturing the main features of the dark matter distribution while highlighting systematic differences at the center and outskirts that may point to galaxy-specific physical conditions or departures from the idealized universal NFW halo model.}

\section{Conclusion and Discussion}
\label{sec:conclusion}

{
In this work, we have carried out a comprehensive general relativistic investigation of the spiral galaxy NGC~7331 by combining high-quality rotation curve data with independent photometric constraints on the stellar mass distribution derived from WISE W1 (3.4~$\mu$m) observations. 

Adopting a static, spherically symmetric spacetime geometry supported by an anisotropic matter distribution with vanishing radial pressure, we reconstructed the galactic gravitational structure directly from the observed kinematics.

A modified exponential form was shown to provide an excellent description of the observed rotation curve across the full radial extent of the galaxy. The robustness of this fit was confirmed through extensive statistical validation, including goodness-of-fit tests, residual analysis, and cross-validation schemes. 

Using the fitted velocity profile, we derived explicit expressions for the redshift function, enclosed mass, energy density, and tangential pressure within the relativistic framework.

The stellar mass profile obtained from WISE photometry exhibits a rapid rise in the inner regions followed by saturation at large radii, consistent with an extended stellar disk dominated by old stellar populations. 

A direct comparison between the photometric stellar mass and the total gravitating mass inferred from the rotation curve reveals a systematic excess mass beyond the optical disk, clearly indicating the presence of a dominant non-luminous component. This excess mass is naturally interpreted as a dark matter halo.

The physical viability of the resulting relativistic halo model was carefully examined. All standard energy conditions---Null, Weak, Strong, and Dominant---are satisfied throughout the halo, while causality is preserved as the squared sound speed remains subluminal across the entire radial range. 

An analysis of the effective potential confirms the existence of stable circular orbits, further supporting the dynamical consistency of the model. The gravitational energy of the system is found to be negative, demonstrating that the halo configuration is gravitationally bound and attractive.

The effective pressure--density relation and the corresponding equation-of-state parameter remain small and positive, indicating that pressure contributions are subdominant compared to the energy density. This behavior is consistent with expectations for dark-matter-dominated systems on galactic scales and supports the interpretation of the halo matter as weakly pressured and anisotropic within a general relativistic setting.

A comparison with the standard Navarro--Frenk--White (NFW) profile shows good agreement at intermediate radii, where observational constraints from the rotation curve are strongest. Deviations in the inner and outer regions---characterized by a flatter central density and a slower decline at large radii---suggest that the halo of NGC~7331 may depart from the idealized universal NFW form, possibly reflecting galaxy-specific formation history or baryonic effects.

Overall, this study demonstrates that the mass distribution and gravitational structure of NGC~7331 can be consistently interpreted within a general relativistic framework when rotation curve data are combined with independent photometric stellar mass estimates. The results highlight the usefulness of relativistic modeling as a complementary approach to conventional dark matter halo analyses, offering deeper insight into the interplay between spacetime geometry, anisotropic stresses, and galactic dynamics.}

\section{Data Availability}

The data underlying this article will be shared on reasonable request to the corresponding author.

\section*{Acknowledgements}

The authors sincerely thank the referee for a careful reading of the manuscript and for the constructive comments and suggestions, which have significantly improved the clarity and quality of this work. FR gratefully acknowledges support from RUSA~2.0 and the Anusandhan National Research Foundation (ANRF). FR, BSC, and AS are thankful to the Inter-University Centre for Astronomy and Astrophysics (IUCAA), Pune, India, for their support. FR, BSC, and AS also gratefully acknowledge academic support from Jadavpur University (JU). BSC acknowledges financial assistance from the University Grants Commission (UGC). FR further thanks ANRF and the Science and Engineering Research Board (SERB), Government of India, for their support.

\nocite{*}
\bibliographystyle{mnras}
\bibliography{refer}

\end{document}